\documentclass[twocolumn,showpacs,showkeys,preprintnumbers,amsmath,amssymb]{revtex4}
\usepackage{graphicx}% Include figure files
\usepackage{dcolumn}% Align table columns on decimal point
\usepackage{bm}% bold math

\begin{document}

\preprint{APS/123-QED}

\title{Exact solution of the mixed-spin Ising model on a decorated square lattice with  
       two different kinds of decorating spins on horizontal and vertical bonds}
\author{Jozef Stre\v{c}ka} 
\email{jozef.strecka@upjs.sk, jozkos@pobox.sk} 
\homepage{http://158.197.33.91/~strecka}
\author{Lucia \v{C}anov\'a}
\author{Michal Ja\v{s}\v{c}ur}
\affiliation{Department of Theoretical Physics and Astrophysics, 
Faculty of Science, \\ P. J. \v{S}af\'{a}rik University, Park Angelinum 9,
040 01 Ko\v{s}ice, Slovak Republic}

\date{\today}
             
\begin{abstract}
The mixed spin-(1/2, $S_{\rm B}$, $S_{\rm C}$) Ising model on a decorated square lattice with two different kinds of decorating spins $S_{\rm B}$ and $S_{\rm C}$ ($S_{\rm B} \neq S_{\rm C}$) placed on its horizontal and vertical bonds, respectively, is exactly solved by establishing a precise mapping relationship with the corresponding spin-1/2 Ising model on an anisotropic square (rectangular) lattice. The effect of uniaxial single-ion anisotropy acting on both types of decorating spins $S_{\rm B}$ and $S_{\rm C}$ is examined in particular. If decorating spins $S_{\rm B}$ and $S_{\rm C}$ are integer and half-odd-integer, respectively, 
or if the reverse is the case, the model under investigation displays a very peculiar critical behavior 
beared on the spontaneously ordered 'quasi-1D' spin system, which appears as a result of the single-ion anisotropy strengthening. We have found convincing evidence that this remarkable spontaneous ordering virtually arises even though all integer-valued decorating spins tend towards their 'non-magnetic' spin state $S=0$ and the system becomes disordered only upon further increase of the single-ion anisotropy. The single-ion anisotropy parameter is also at an origin of various temperature dependences of the total magnetization when imposing the pure ferrimagnetic or the mixed ferro-ferrimagnetic character of the spin arrangement.
\end{abstract}

\pacs{05.50.+q, 05.70.Jk, 64.60.Cn, 75.10.Hk, 75.10.-b, 75.30.Kz, 75.40.Cx} 
\keywords{Ising model, mixed spins, critical behavior, single-ion anisotropy, exact results}

\maketitle

\section{\label{sec:intro}Introduction}

One of the most fascinating and challenging topics in equilibrium statistical physics are phase transitions and critical phenomena of rigorously solvable interacting many-body systems \cite{baxt82}. The {\it planar Ising model} represents perhaps the simplest lattice-statistical model for which a complete exact solution 
is known since Onsager's famous solution \cite{onsa44} and the model simultaneously undergoes a non-trivial 
phase transition. Although the spin-1/2 Ising model on two-dimensional (2D) lattices with non-crossing bonds 
is in principle exactly soluble problem within the Pfaffian method \cite{gree64}, its precise treatment 
usually requires an application of sophisticated mathematical methods, which consequently lead to 
considerable difficulties when applying them to more complex models describing spin systems with 
interactions beyond nearest neighbors \cite{nnn}, frustrated spin systems \cite{frustr}, 
or higher-spin models with or without single-ion anisotropy and biquadratic interactions \cite{higher}. 
It is noteworthy, however, that exact solutions of the spin-1/2 Ising model have already been 
reported on several Archimedean lattices (square \cite{onsa44}, triangular and honeycomb \cite{triang}, kagom\'e \cite{syoz51}, extended kagom\'e \cite{312}, bathroom-tile \cite{48}, orthogonal-dimer \cite{stre05} and ruby \cite{lin83} lattice) and more recently, also on a variety of more complex irregular 
lattices such as union jack \cite{union}, pentagonal \cite{urum02}, square-kagom\'e \cite{sun06}, 
as well as, two topologically different square-hexagonal \cite{46} lattices. An importance of the 
rigorously solved 2D Ising lattices can be viewed in providing guidance on the scaling 
and universality hypotheses \cite{scaleuniv} and moreover, these exact results provide very valuable information about an accuracy of different approximative theories used to study spin systems where 
the rigorous treatment is inapplicable. Even though the Ising model has been originally designed 
for describing essential features of insulating magnetic materials \cite{dejo74}, throughout the years 
various modifications of this model have found rich applications in seemingly diverse research areas 
\cite{dick00}. 

Over the last few decades, the mixed-spin Ising models have attracted a great deal of research interest 
on behalf of much richer critical behavior they display compared with their single-spin counterparts. Actually, the mixed-spin Ising models are often convenient candidates for displaying tricritical 
phenomenon or other complicated change of an usual critical point to a multicritical point. 
Despite much efforts, there are only few {\it exactly solvable mixed-spin Ising models}, yet.
Using the generalized decoration-iteration and star-triangle mapping transformations, the mixed 
spin-(1/2, $S$) Ising model on the honeycomb, diced and several decorated planar lattices have exactly 
been treated by Fisher and Yamada many years ago \cite{genhigh}. Notice furthermore that an extension 
of the generalized mapping transformations enabled to account also for the single-ion anisotropy effect. 
The effect of uniaxial and biaxial single-ion anisotropies has been subsequently exactly examined in 
the mixed-spin Ising models on three-coordinated honeycomb  \cite{gonca} and bathroom-tile \cite{stre06} lattices, diced lattice \cite{jasc05} and some decorated lattices \cite{jasc98,oitm05}. To the best of our knowledge, these are the only mixed-spin planar Ising models with generally known exact solutions except several mixed-spin Ising models on the Bethe lattices studied within the framework of exact recursion relations \cite{bethe}. Among the striking models for which a precise solution is restricted to a certain subspace of interaction parameters only, one should further mention the mixed-spin Ising model on the union 
jack (centered square) lattice, which can be mapped onto a symmetric (zero-field) eight-vertex 
model with continuously varying critical exponents \cite{lipo95}.  

Exactly solvable {\it Ising models on 2D decorated lattices}, the bonds of which are decorated 
in various fashion by additional spins, are therefore of particular research interest 
(see Ref.~\onlinecite{syoz72} and references cited therein). In the class of exactly solved 
decorated Ising models belong the original ferrimagnetic model introduced by Syozi and Nakano \cite{ferri}, partly \cite{partly} and multiply \cite{multi} decorated models showing reentrant phase transitions, 
ANNNI models \cite{huse81}, diluted models of ferromagnetism \cite{dilute}, decorated models 
with classical $\nu$-dimensional vector spins \cite{vector}, Fisher's superexchange models and its numerous variants \cite{super}, as well as, the models with higher decorating spins \cite{genhigh}. It is worthwhile 
to remark that rigorous solutions of these models have furnished answers to questions interesting both 
from the academic point of view (scaling and universality hypotheses, reentrant phase transitions), 
as well as, from the experimental point of view (dilution, technological applications of ferrimagnets). 
The vast potential of ferrimagnets with respect to technological applications has also stimulated 
exploration of the effect of single-ion anisotropy upon ferrimagnetic features of the mixed spin-(1/2, $S$) Ising models on the wholly \cite{jasc98} and partly \cite{oitm05} decorated lattices. 

Recently, Kaneyoshi \cite{kane01} has proposed another notable example of the decorated Ising model 
on a square lattice the horizontal and vertical bonds of which are occupied by two different kinds 
of decorating spins. Up to now, this remarkable model system have been studied using the approach based 
on the differential operator technique \cite{honm79}, whereas an accuracy of obtained results determined 
the Bethe-Peierls-Weiss approximation used for the undecorated lattice \cite{kane01}. Therefore, 
the main aim of the present work is to extend the class of exactly solved Ising models by providing 
an accurate solution for this mixed spin-(1/2, $S_{\rm B}$, $S_{\rm C}$) Ising model on the square lattice
with two different kinds of decorating spins $S_{\rm B}$ and $S_{\rm C}$ ($S_{\rm B} \neq S_{\rm C}$). 
Exact results for the system under consideration are obtained by applying the generalized decoration-iteration transformation \cite{syoz72} establishing an exact mapping correspondence with an effective spin-1/2 Ising model on the anisotropic square (rectangular) lattice. Owing to this fact, the known exact solution of the latter spin-1/2 Ising model on the rectangular lattice \cite{onsa44} straightforwardly enables to acquire exact results for the former decorated mixed-spin model. Within the framework of this exact method, 
we will concentrate our attention first of all to the influence of the single-ion anisotropy on the 
critical behavior and phase diagrams. Besides, temperature dependences of the total magnetization will be also particularly examined. 

The outline of this paper is as follows. In Section \ref{sec:model}, the detailed description of the considered model system is presented in the first instance. Then, some details of the mapping method will be clarified  along with the derivation of exact expressions for the magnetization and critical temperatures. 
The most interesting results are presented and detailed discussed in Section \ref{sec:result} 
for two particular sets of the quantum spin numbers ($S_{\rm A}$, $S_{\rm B}$, $S_{\rm C}$) = 
(1/2, 1, 3/2) and (1/2, 2, 3/2). Finally, some concluding remarks are mentioned in Section \ref{sec:conc}. 

\section{\label{sec:model}Model and method}

\begin{figure}
\begin{center}
\includegraphics[width=7cm]{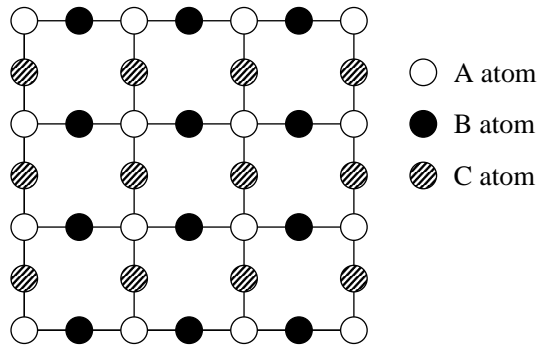} 
\end{center} 
\vspace{-4mm} 
\caption{A cross-section of the three sublattice (ternary) mixed-spin Ising model on the decorated 
square lattice. Open circles denote lattice positions of the spin-1/2 atoms (sublattice A), while black 
and hatched circles label lattice positions of the decorating spin-$S_{\rm B}$ atoms (sublattice B) 
and the spin-$S_{\rm C}$ atoms (sublattice C), respectively.}
\label{fig1}
\end{figure}

Suppose the square lattice with two different kinds of decorating spins $S_{\rm B}$ and $S_{\rm C}$ placed 
on its horizontal and vertical bonds, respectively, as it is diagrammatically depicted in FIG.~\ref{fig1}. 
As it can be clearly seen, the displayed magnetic structure can be also viewed as the three sublattice (ternary) mixed-spin system in which each vertex of the original square lattice is occupied by the spin-$S_{\rm A}$ atom (sublattice A), while each its horizontal and vertical bond is decorated by the spin-$S_{\rm B}$ (sublattice B) and spin-$S_{\rm C}$ (sublattice C) atom, respectively. As a result, the nearest-neighbor Ising model defined upon the underlying lattice is given by the Hamiltonian
\begin{eqnarray}
{\cal H} &=& J_{\rm AB} \! \! \! \! \! \!  \! \sum_{(i,j) \in horiz.}  \! \! \! \! \! \! \!  
             S^{\rm A}_{i} S^{\rm B}_{j}
          + J_{\rm AC}  \! \! \! \! \! \! \sum_{(i,k) \in vert.}  \! \! \! \! \! \! 
            S^{\rm A}_{i} S^{\rm C}_{k} \nonumber \\
         &-& D_{\rm B} \sum_{j = 1}^{N} (S^{\rm B}_{j})^2 - D_{\rm C} \sum_{k = 1}^{N} (S^{\rm C}_{k})^2,
\label{eq:H}
\end{eqnarray}
where $S_{i}^{\rm A} = \pm \frac{1}{2}$, $S^{\rm B}_{j} = -S_{\rm B}, -S_{\rm B}+1, \dots, S_{\rm B}$ and 
$S^{\rm C}_{k} = -S_{\rm C}, -S_{\rm C}+1, \dots, S_{\rm C}$ denote three different kinds of Ising spins 
differing in value, their superscript determines the sublattice to which spins belong, whereas  
their subscript specifies lattice points where the spins are placed. The first two summations 
are carried out over the nearest-neighbor A--B and A--C spin pairs, respectively, and the 
last two summations are taken over all decorating B and C sites. Accordingly, the parameters $J_{\rm AB}$ 
and $J_{\rm AC}$ label pairwise exchange interactions between the nearest-neighbor A--B and A--C spin pairs, while the single-spin parameters $D_{\rm B}$ and $D_{\rm C}$ measure a strength of the uniaxial single-ion anisotropy acting on decorating spins $S_{\rm B}$ and $S_{\rm C}$, respectively. The essence of both 
single-ion anisotropy parameters $D_{\rm B}$ and $D_{\rm C}$ lies in an uniaxial magnetic anisotropy, which comes from a crystal field of ligands from a local neighborhood of magnetic centers \cite{boca04}.

The crucial step in our approach represents evaluation of the partition function, i.e. a sum over all possible spin configurations, what usually represents a formidable mathematical task in a highly cooperative spin system with many microscopical degrees of freedom. Fortunately, the problem connected with calculation 
of the partition function of decorated models can be simply avoided by the use of well-known trick, 
which consists in performing the summation over spin degrees of freedom of all {\it decorating spins} 
$S_{\rm B}$ and $S_{\rm C}$ before summing over spin degrees of freedom of the {\it vertex spins} $S_{\rm A}$. 
In doing so, the partition function can be partially factorized into the following product
\begin{eqnarray}
{\cal Z} = \sum_{\{S^{\rm A}_{i} \}} \prod_{j=1}^{N} \! W_{\rm B} (S^{\rm A}_{j1} + S^{\rm A}_{j2})
                                     \prod_{k=1}^{N} \! W_{\rm C} (S^{\rm A}_{k1} + S^{\rm A}_{k2}), 
\label{eq:Z}
\end{eqnarray}
where the symbol $\sum_{\{S^{\rm A}_{i} \}}$ stands for the summation over all available spin configurations on the sublattice A and the first (second) product is performed over all decorating spins $S_{\rm B}$ ($S_{\rm C}$) occupying the horizontal (vertical) bonds. Furthermore, the function $W_{\Omega}(x)$, which depends on two vertex spins $S^{\rm A}_{i1}$ and $S^{\rm A}_{i2}$ from the sublattice A coupled indirectly via $i$th decorating spin, marks the expression
\begin{eqnarray}
W_{\Omega}(x)= \sum_{n=-S_{\Omega}}^{+S_{\Omega}} \exp(\beta D_{\Omega} n^2) 
               \cosh(\beta J_{{\rm A} \Omega} n x), 
\label{eq:W}
\end{eqnarray}
where $\Omega$ = B or C, $\beta = 1/(k_{\rm B}T)$, $k_{\rm B}$ is being Boltzmann's constant 
and $T$ is the absolute temperature.

In order to proceed further with calculation, the generalized decoration-iteration mapping 
transformation  \cite{syoz72} can be now employed
\begin{eqnarray}
W_{\Omega} (S^{\rm A}_{i1} + S^{\rm A}_{i2}) = A_{\Omega} \exp(\beta R_{\Omega} S^{\rm A}_{i1} 
S^{\rm A}_{i2}), \hspace{0.2cm} \Omega = {\rm B \, or \, C}.
\label{eq:DIT}
\end{eqnarray}
The physical meaning of the mapping (\ref{eq:DIT}) is to remove all interaction parameters associated 
with one decorating spin ($S_{\rm B}$ or $S_{\rm C}$) and to replace them by a new effective interaction
($R_{\rm B}$ or $R_{\rm C}$) between the remaining vertex spins $S^{\rm A}_{i1}$ and $S^{\rm A}_{i2}$. 
It is worthwhile to remark that a self-consistency condition of the mapping relation (\ref{eq:DIT})  unambiguously determines both unknown mapping parameters $A_{\Omega}$ and $R_{\Omega}$, since it 
must hold independently of the spin states of both vertex spins included in this transformation. 
As a matter of fact, the direct substitution of four possible spin combinations of two vertex spins 
$S^{\rm A}_{i1}$ and $S^{\rm A}_{i2}$ indeed gives just two independent equations from the formula (\ref{eq:DIT}), which subsequently unambiguously determine the mapping parameters $A_{\Omega}$ and $R_{\Omega}$ 
\begin{eqnarray}
A_{\Omega} &=& [W_{\Omega}(1) W_{\Omega}(0)]^{\frac{1}{2}}, \hspace{0.9cm} \Omega = {\rm B \, or \, C}, \label{eq:A} \\
\beta R_{\Omega} &=& 2 \ln[W_{\Omega}(1)/W_{\Omega}(0)], \hspace{0.4cm}    \Omega = {\rm B \, or \, C}.
\label{eq:R}
\end{eqnarray}
At this stage, let us substitute the transformation (\ref{eq:DIT}) into the expression (\ref{eq:Z}) 
in order to gain the relation
\begin{equation}
{\cal Z} (\beta, J_{\rm AB}, J_{\rm AC}, D_{\rm B}, D_{\rm C}) 
= A_{\rm B}^N A_{\rm C}^N {\cal Z}_{0}(\beta R_{\rm B}, \beta R_{\rm C}),
\label{eq:MR}
\end{equation}
which relates the partition function ${\cal Z}$ of the mixed-spin Ising model on the decorated lattice 
with the partition function ${\cal{Z}}_{0}$ of the undecorated spin-$1/2$ Ising model on the anisotropic square (rectangular) lattice with two different nearest-neighbor couplings $R_{\rm B}$ and $R_{\rm C}$ in the horizontal and vertical directions, respectively. Nevertheless, it should be mentioned here that the 
mapping relation (\ref{eq:MR}) between both the partition functions represents a central result of our calculation, since it formally completes an exact solution of the partition function ${\cal Z}$ with 
regard to the known exact result for the partition function ${\cal Z}_{0}$ of the spin-1/2 Ising model 
on the rectangular lattice \cite{onsa44} given by the effective Hamiltonian 
${\cal H}_0 = - R_{\rm B} \sum_{(i,j)}^{horiz.} S^{\rm A}_{i} S^{\rm A}_{j} 
              - R_{\rm C} \sum_{(k,l)}^{vert.} S^{\rm A}_{k} S^{\rm A}_{l}$. 

Now, we turn to calculation of the sublattice magnetization. With the help of commonly used mapping theorems \cite{barr88}, one easily proves a validity of the following relation for the spontaneous magnetization $m_{\rm A}$ of the sublattice A
\begin{equation}
m_{\rm A} \equiv \langle S_i^{\rm A} \rangle = \langle S_i^{\rm A} \rangle_0 \equiv m_{0},
\label{eq:MA}
\end{equation}
where the symbols $\langle \ldots \rangle$ and $\langle \ldots \rangle_0$ denote canonical ensemble averaging performed within the decorated and its corresponding undecorated model system, respectively. Apparently, the sublattice magnetization $m_{\rm A}$ directly equals to the magnetization $m_{0}$ of the corresponding spin-1/2 Ising model on the rectangular lattice with the effective horizontal and vertical coupling constants given by the equation (\ref{eq:R}). When taking into account the Potts's and Chang's \cite{pott52} exact result for the spontaneous magnetization of the spin-1/2 Ising model on the rectangular lattice, it is possible to write the following expression for the sublattice magnetization 
$m_{\rm A}$
\begin{eqnarray}
m_{\rm A} = \frac{1}{2} \Bigl[1 - \sinh^{-2} \Bigl(\frac{\beta R_{\rm B}}{2}\Bigr) 
                                  \sinh^{-2} \Bigl(\frac{\beta R_{\rm C}}{2}\Bigr) \Bigr]^{\frac{1}{8}}, 
\label{eq:MAF1}
\end{eqnarray}
or equivalently,
\begin{eqnarray}
m_{\rm A} = \frac{1}{2} \biggl \{ 1 - \frac{16 W_{\rm B}^2 (1) W_{\rm B}^2(0) W_{\rm C}^2 (1) W_{\rm C}^2(0)}
                               {[W_{\rm B}^2 (1) - W_{\rm B}^2(0)]^2 [W_{\rm C}^2 (1)- W_{\rm C}^2(0)]^2}                            \biggr \}^{\frac{1}{8}}.
\label{eq:MAF2}
\end{eqnarray}
Next, it is also of particular interest to derive exact expressions for the spontaneous magnetization of 
the sublattice B and C constituted by the higher decorating spins $S_{\rm B}$ and $S_{\rm C}$, respectively.
For this purpose, the exact Callen-Suzuki \cite{call63} spin identity formulated generally for both kinds 
of decorating spins can be successfully utilized 
\begin{equation}
m_{\Omega} = \langle S_i^{\Omega} \rangle  = 
\biggl \langle \frac{\sum_{S_i^{\Omega}} S_i^{\Omega} \exp(-\beta {\cal H}_i)}
                   {\sum_{S_i^{\Omega}} \exp(-\beta {\cal H}_i)} 
\biggr \rangle,      
\label{eq:CS}
\end{equation}
in which $\Omega$ = B or C and the Hamiltonian ${\cal H}_i$ labels such a part of the total Hamiltonian (\ref{eq:H}), which contains all interaction terms of one decorating spin $S_i^{\Omega}$. By making use 
of the mapping relation (\ref{eq:MR}) and the exact spin identity (\ref{eq:CS}), it can be easily proved 
that the sublattice magnetization $m_{\rm B}$ and $m_{\rm C}$ satisfy the equality 
\begin{equation}
m_{\Omega} = \langle F_{\Omega} (S_{i1}^{\rm A} + S_{i1}^{\rm A}) \rangle,   
             \hspace{0.6cm}    \Omega = {\rm B \, or \, C},
\label{eq:CS1}
\end{equation}
where the function $F_{\Omega} (x)$ is defined as follows
\begin{equation}
F_{\Omega} (x) = - \frac{\displaystyle 
   \sum_{n=-S_{\Omega}}^{S_{\Omega}} n \exp(\beta D_{\Omega} n^2) \sinh(\beta J_{\rm A \Omega} n x)}
{\displaystyle \sum_{n=-S_{\Omega}}^{S_{\Omega}} \exp(\beta D_{\Omega} n^2) \cosh(\beta J_{\rm A \Omega} n x)}.
\label{eq:FO}
\end{equation}
By employing the differential operator technique \cite{honm79} together with the exact van der Waerden identity $\exp(a S^{{\rm A}}_i) = \cosh(a/2) + 2S^{{\rm A}}_i \sinh(a/2)$ into the equation (\ref{eq:CS1}), 
one readily finds that both the sublattice magnetization $m_{\rm B}$ and $m_{\rm C}$ can be expressed 
solely as a function of the sublattice magnetization $m_{\rm A}$, in fact,  
\begin{equation}
m_{\Omega} = \langle \exp[(S_{i1}^{\rm A} + S_{i1}^{\rm A}){\rm d}/{\rm d}x] \rangle F_{\Omega} (x) 
           = 2 m_{\rm A} F_{\Omega} (1).
\label{eq:MBC}
\end{equation}
The above result completes an exact solution for the sublattice magnetization $m_{\rm B}$ and $m_{\rm C}$ 
in view of the formerly derived exact result (\ref{eq:MAF1}) [or (\ref{eq:MAF2})] for the sublattice magnetization $m_{\rm A}$.

Finally, let us turn our attention to the critical behavior of the model under investigation. It is quite
obvious from the equation (\ref{eq:MBC}) that all sublattice magnetization tend necessarily to zero if and only if the sublattice magnetization $m_{\rm A}$ goes to zero. Accordingly, the critical temperature can be located from the condition which is consistent with the Onsager's critical condition for the spin-1/2 Ising model on the rectangular lattice \cite{onsa44}
\begin{equation}
\sinh(\beta_{\rm c} R_{\rm B}/2) \sinh(\beta_{\rm c} R_{\rm C}/2) = 1,
\label{eq:Tc1}
\end{equation}
where $\beta_{\rm c} = 1/(k_{\rm B}T_{\rm c})$ is being defined as the inverse critical temperature in energy units, $T_{\rm c}$ denotes the critical temperature and the effective coupling constants $R_{\rm B}$ and $R_{\rm C}$ are given by the equation (\ref{eq:R}). It is also interesting to mention that the critical temperature can be obtained from the alternate condition, which is of course equivalent to the condition (\ref{eq:Tc1}), but without referring to the effective coupling constants of the undecorated rectangular lattice. The expression (\ref{eq:MAF2}) for the sublattice magnetization $m_{\rm A}$ indeed yields the 
following critical condition
\begin{eqnarray}
\{ [W_{\rm B}^{\rm c}(1)]^2 \! \! &-& \! \! [W_{\rm B}^{\rm c}(0)]^2 \} 
\{ [W_{\rm C}^{\rm c}(1)]^2 - [W_{\rm C}^{\rm c}(0)]^2 \} \nonumber \\
&=& 4 W_{\rm B}^{\rm c} (1) W_{\rm B}^{\rm c} (0) W_{\rm C}^{\rm c} (1) W_{\rm C}^{\rm c} (0).
\label{eq:Tc2}
\end{eqnarray}
The superscript in the aforementioned expressions means that the inverse critical temperature 
$\beta_{\rm c}$ enters into the relevant expressions (\ref{eq:W}) instead of $\beta$. 

\section{\label{sec:result}Results and discussion}

Before proceeding to a discussion of the most interesting results, let us make few remarks on a validity 
of the results to be obtained in the preceding section. Notice initially that the obtained results are 
rather general, actually, they hold regardless of whether ferromagnetic or antiferromagnetic interactions 
$J_{\rm AB}$ and $J_{\rm AC}$ are considered, irrespective of the value of the decorating spins 
($S_{\rm B}$ and $S_{\rm C}$) and even both the single-ion anisotropy parameters $D_{\rm B}$ and $D_{\rm C}$ 
can be taken independently of each other. It is noteworthy, however, that several studies reported on 
previously have already involved some particular cases of the model under investigation. By imposing 
$S_{\rm B}$ = $S_{\rm C}$, $J_{\rm AB} = J_{\rm AC}$ and $D_{\rm B}$ = $D_{\rm C}$, for instance, our 
results reduce to those acquired for the mixed-spin Ising model on a symmetrically decorated square lattice adapted to study essential features of the ferrimagnetism \cite{jasc98}. In the present paper, we will therefore restrict our attention only to the particular case with two different kinds of decorating spins $S_{\rm B} \neq S_{\rm C}$. More specifically, one of the two decorating spins (say $S_{\rm B}$) is 
assumed to be integer valued, while the other one (say $S_{\rm C}$) is anticipated to be half-odd-integer. 
For simplicity, another constraint introduced through an equality between the single-ion anisotropy 
terms $D_{\rm B} = D_{\rm C} = D$ will be supposed in order to reduce the number of free parameters involved in the model Hamiltonian (\ref{eq:H}). Other particular models in which both the decorating spins 
are integer or half-odd-integer, respectively, will be explored in the subsequent separate works \cite{cano06a,cano06b}.

\begin{figure*}
\begin{center}
\includegraphics[width=14cm]{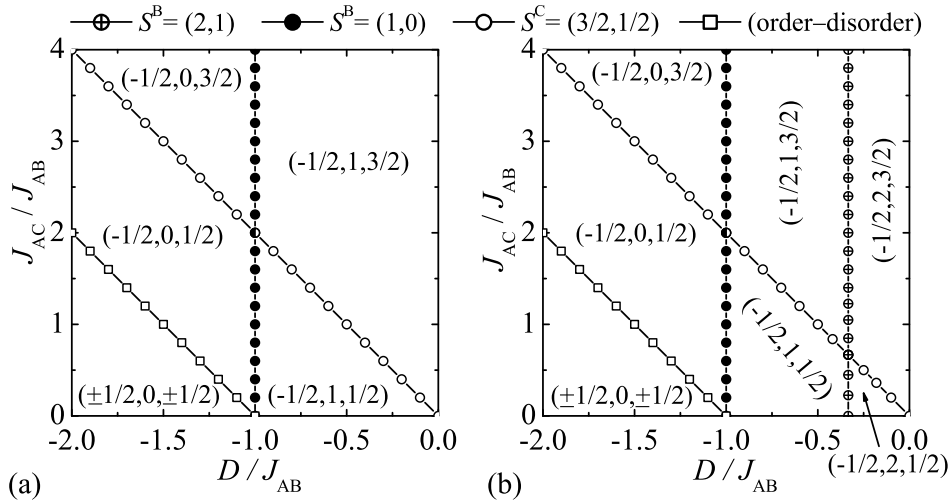} 
\end{center} 
\vspace{-12mm} 
\caption{Ground-state phase diagrams in the $D/J_{\rm AB}-J_{\rm AC}/J_{\rm AB}$ plane for two 
particular spin cases with $(S_{\rm A}, S_{\rm B}, S_{\rm C})=(1/2, 1, 3/2)$ [FIG.~\ref{fig2}(a)] 
and $(S_{\rm A}, S_{\rm B}, S_{\rm C})=(1/2, 2, 3/2)$ [FIG.~\ref{fig2}(b)]. Round brackets indicate 
spin order to emerge within different sectors of the phase diagram of the pure ferrimagnetic model 
with $J_{\rm AB}>0$ and $J_{\rm AC}>0$. Different symbols to be defined in the legend characterize 
a change in the spin state that occurs at the displayed phase boundaries. Note that a sign change 
in $J_{\rm AB}$ ($J_{\rm AC}$) leads merely to a sign change in the spin state of the spin-$S_{\rm B}$ 
($S_{\rm C}$) atoms. The hollow-square line represents transitions between the ordered and disordered
phases (for details see the text).}
\label{fig2}
\end{figure*}

First, let us take a closer look at the ground-state behavior. In FIG.~\ref{fig2}, the ground-state phase diagrams in the $D/J_{\rm AB}-J_{\rm AC}/J_{\rm AB}$ plane are depicted for two particular models with $(S_{\rm A}, S_{\rm B}, S_{\rm C})=(1/2, 1, 3/2)$ and $(S_{\rm A}, S_{\rm B}, S_{\rm C})=(1/2,2,3/2)$.
As one can see, the interaction parameters $J_{\rm AB}$ and $J_{\rm AC}$ energetically favor the highest 
spin state of the decorating spins $S_{\rm B}$ and $S_{\rm C}$, respectively. Contrary to this, the easy-plane single-ion anisotropy ($D<0$) has a tendency to lower the spin states of the decorating atoms. An eventual spin arrangement is thus determined by a mutual competition between the exchange interactions and the single-ion anisotropy. It is quite obvious from FIG.~\ref{fig2} that there is considerable similarity between both the displayed phase diagrams. As a matter of fact, the only difference consists in the phase boundary that accompanies the transition $2 \leftrightarrow 1$ of the spin-$S_{\rm B}$ atoms, which is of course missing in the first phase diagram. Finally, it is worthy to note the general condition allocating ground-state phase boundaries associated with the spin change $S^{\Omega} \leftrightarrow S^{\Omega} - 1$ 
\begin{eqnarray}
\frac{D_{S^{\Omega}, S^{\Omega} - 1}}{J_{\rm A \Omega}} = \frac{1}{1 - 2 S^{\Omega}}, 
\hspace{0.6cm} \Omega = {\rm B \, or \, C}.
\label{eq:pb}	
\end{eqnarray}

\begin{figure*}
\begin{center}
\includegraphics[width=14cm]{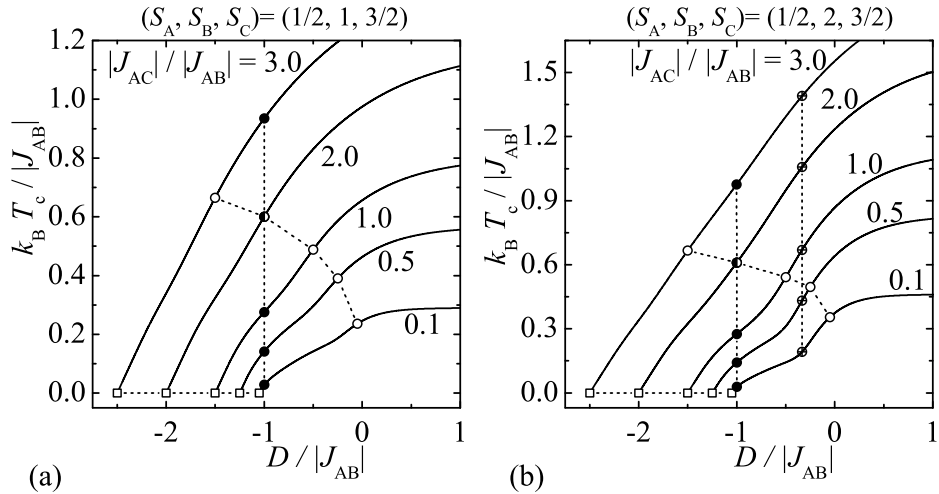} 
\end{center} 
\vspace{-12mm} 
\caption{Critical temperature as a function of the single-ion anisotropy for several values of 
the ratio $|J_{\rm AC}|/|J_{\rm AB}|$ in two investigated spin systems with $(S_{\rm A}, S_{\rm B}, 
S_{\rm C})=(1/2, 1, 3/2)$ [FIG.~\ref{fig3}(a)] and $(S_{\rm A}, S_{\rm B}, S_{\rm C})=(1/2, 2, 3/2)$ [FIG.~\ref{fig3}(b)]. Different symbols characterize the same spin transition as explained in the 
legend of the ground-state phase diagram shown in FIG.~\ref{fig2}. Broken lines connecting different 
spin transitions are guides for eyes only.}
\label{fig3}
\end{figure*}

Next, let us investigate in detail the effect of the single-ion anisotropy upon the critical behavior of the considered model system. For this purpose, we have plotted in FIG.~\ref{fig3} the finite-temperature phase diagram in a form of the critical temperature vs. single-ion anisotropy dependence for the two investigated
spin systems. It should be pointed out that 
the unique solution of the critical condition (\ref{eq:Tc1}) [or equivalently (\ref{eq:Tc2})] gives 
all the displayed phase boundaries, which consequently represent the lines of second-order (continuous) transitions between the spontaneously ordered phases and disordered (paramagnetic) phase. Furthermore, 
it is worth noticing that the interaction parameters $J_{\rm AB}$ and $J_{\rm AC}$ enter within the 
critical conditions (\ref{eq:Tc1}) and (\ref{eq:Tc2}) only into the arguments of even functions. 
In this regard, the depicted phase boundaries remain in force irrespective of a sign of the 
interaction parameters $J_{\rm AB}$ and $J_{\rm AC}$, what means that the same phase boundaries 
hold regardless of whether the pure ferrimagnetic, the mixed ferro-ferrimagnetic, or 
even the pure ferromagnetic system is considered. 

The most striking finding resulting from the finite-temperature phase diagram shown in FIG.~\ref{fig3} 
is that the lines of phase transitions do not reach zero temperature at the boundary value of the 
single-ion anisotropy $D_{1,0}/J_{\rm AB} = -1$, which is sufficiently strong to bring all the 
integer-valued decorating spins $S_{\rm B}$ into their 'non-magnetic' spin state $S^{\rm B} = 0$. 
Namely, one would intuitively expect that all the phases with the 'non-magnetic' decorating spins 
$S_{\rm B}$ (i.e. the phases appearing in FIG.~\ref{fig2} in the parameter space where $D<D_{1,0}$) 
should be in fact disordered due to their 'quasi-1D' character. 
In the zero-temperature limit, one actually finds that the effective coupling constant $\beta R_{\rm B}$ substituting the integer-valued decorating spins $S_{\rm B}$ tends to zero, however, the effective coupling constant $\beta R_{\rm C}$ substituting the half-odd-integer decorating spins $S_{\rm C}$ tends to infinity. In the consequence of that, the 2D decorated mixed-spin system indeed behaves as the 'quasi-1D' spin system, since it effectively splits into a set of the independent mixed spin-($S_{\rm A}$, $S_{\rm C}$) Ising chains. On the other hand, one should bear in mind that the system is spontaneously long-range 
ordered if and only if $\sinh(\beta R_{\rm B}/2) \sinh(\beta R_{\rm C}/2) > 1$, while it becomes disordered 
just as $\sinh(\beta R_{\rm B}/2) \sinh(\beta R_{\rm C}/2) < 1$ in accord with the critical 
condition (\ref{eq:Tc1}). In the limit of zero temperature, it can be easily verified that 
\begin{eqnarray*}
\lim_{T \to 0} \! \! \! \! \! && \! \! \! \! \! \Bigl [ 
\sinh \Bigl(\frac{\beta R_{\rm B}}{2} \Bigr) \sinh \Bigl( \frac{\beta R_{\rm C}}{2} \Bigr) \Bigr ]
\nonumber \\
&=&  \Bigl \{ \begin{array}{ll} 
\infty  & \, {\rm if} \, \, \, D > - J_{\rm AB} - \frac{1}{2} J_{\rm AC}  \\
 \: 0   & \, {\rm if} \, \, \, D < - J_{\rm AB} - \frac{1}{2} J_{\rm AC} 
         \end{array}  
\end{eqnarray*}
what means that the threshold single-ion anisotropy below which the system becomes disordered (at zero 
as well as any non-zero temperature) is given by the constraint $D_{\rm o-d} = - J_{\rm AB} - \frac{1}{2} J_{\rm AC}$. This non-trivial phase boundary, which cannot be obtained from simple energetic arguments, 
is depicted in the ground-state phase diagram (FIG.~\ref{fig2}) as a hollow-square line. This result 
straightforwardly proves that all the phases appearing above this order-disorder line are at sufficiently 
low temperatures spontaneously long-range ordered in spite of their 'quasi-1D' nature. In agreement with 
the aforementioned arguments, the non-zero critical temperatures to be observed in the parameter space 
where $D < D_{1,0}$ manifest the order-disorder transition between the spontaneously ordered phase 
and the disordered phase even though all the spin-$S_{\rm B}$ atoms reside in the ground 
state of the ordered phase the 'non-magnetic' state $S^{\rm B} = 0$  (see FIG.~\ref{fig3}). 
In this respect, the part of second-order phase transition lines starting at $D = - J_{\rm AB}$ 
and terminating at $D = - J_{\rm AB} - \frac{1}{2} J_{\rm AC}$ (cf. FIGs.~\ref{fig2} and \ref{fig3}) 
can be identified as the critical line at which the spontaneous long-range order of the 'quasi-1D' 
spin systems disappears.

In order to provide an independent check of the aforementioned scenario, the total and sublattice magnetization will be detailed analyzed for the decorated model with $(S_{\rm A}, S_{\rm B}, S_{\rm C})
=(1/2, 1, 3/2)$. Before starting our further analysis, it is worthwhile to mention that this spin system exhibits all generic features of the models with other decorating spin values, as well.   
\begin{figure*}
\begin{center}
\includegraphics[width=14cm]{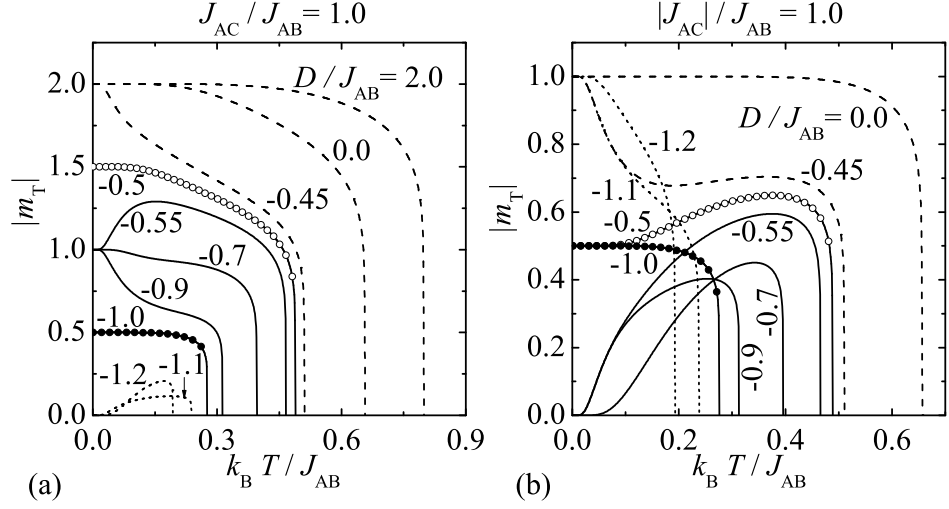} 
\end{center} 
\vspace{-10mm} 
\caption{Temperature dependences of the total magnetization for the decorated mixed-spin system with 
$(S_{\rm A}, S_{\rm B}, S_{\rm C})=(1/2, 1, 3/2)$ when the ratio $|J_{\rm AC}|/J_{\rm AB} = 1.0$ is 
fixed and the single-ion anisotropy strength $D/J_{\rm AB}$ varies. FIG.~\ref{fig4}(a) shows the situation
in the pure ferrimagnetic system with the exchange constants $J_{\rm AB}>0$ and $J_{\rm AC}>0$, whereas FIG.~\ref{fig4}(b) corresponds to the mixed ferro-ferrimagnetic system with $J_{\rm AB}>0$ and 
$J_{\rm AC}<0$. Different line styles correspond to different phases.}
\label{fig4}
\end{figure*}
The total magnetization $m_{\rm T} = |m_{\rm A} +  m_{\rm B} + m_{\rm C}|$ reduced per one atom of the sublattice A is shown in FIG.~\ref{fig4} for the pure ferrimagnetic system with $J_{\rm AB}>0$, 
$J_{\rm AC}>0$ [FIG.~\ref{fig4}(a)] and the mixed ferro-ferrimagnetic system with $J_{\rm AB}>0$, 
$J_{\rm AC}<0$ [FIG.~\ref{fig4}(b)]. Note that different line styles are used to distinguish between 
different spin orderings to emerge in the zero-temperature limit (ground state). Since the sign change 
in $J_{\rm AC}$ leads merely to the sign change in the sublattice magnetization $m_{\rm C}$ ($m_{\rm C}$ 
is an odd function of $J_{\rm AC}$), all temperature variations of the total magnetization can be explained via the same temperature variations of the sublattice magnetization. 

\begin{figure*}
\begin{center}
\includegraphics[width=16cm]{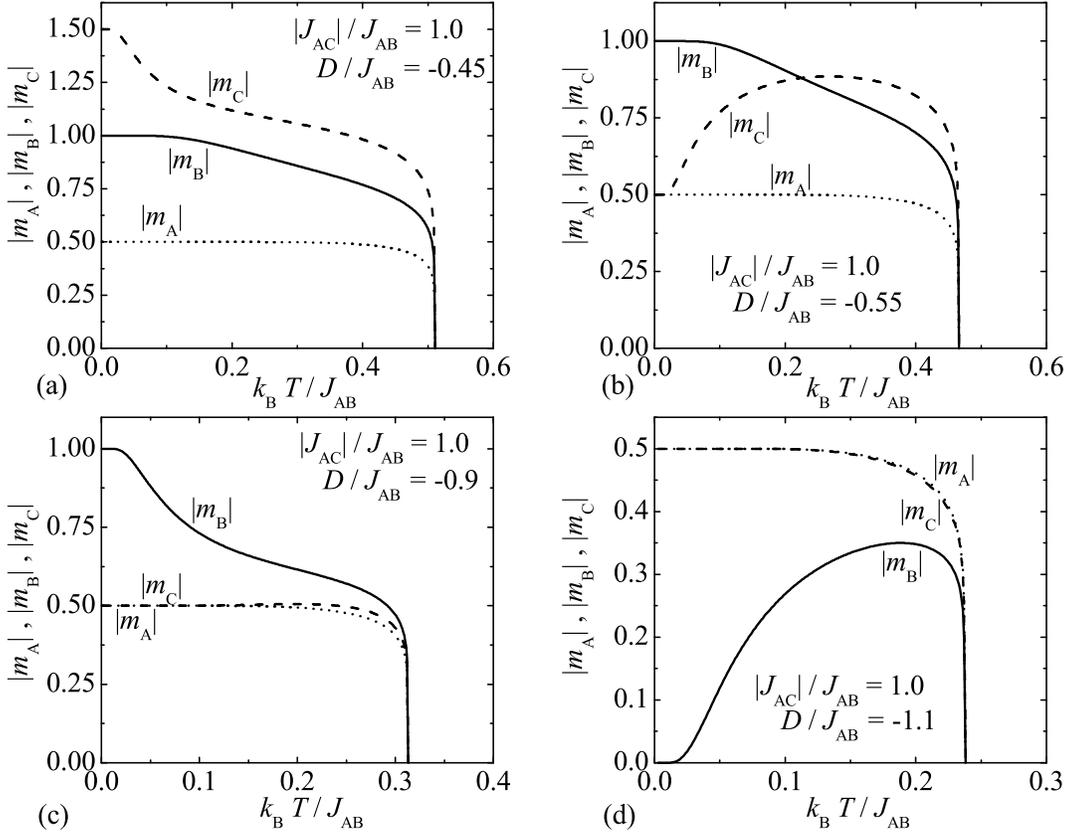} 
\end{center} 
\vspace{-15mm} 
\caption{The absolute value of the sublattice magnetization $m_{\rm A}$ (dotted lines), $m_{\rm B}$ (solid lines), and $m_{\rm C}$ (dashed lines) for $|J_{\rm AC}|/J_{\rm AB} = 1$ and several values of the   single-ion anisotropy parameter $D/J_{\rm AB} = -0.45$ [FIG.~\ref{fig5}(a)], $-0.55$ [FIG.~\ref{fig5}(b)],
$-0.9$ [FIG.~\ref{fig5}(c)], and $-1.1$ [FIG.~\ref{fig5}(d)].}
\label{fig5}
\end{figure*}

For this reason, we depict in FIG.~\ref{fig5} all three sublattice magnetization as a function of the temperature for those particular values of the single-ion anisotropy for which the most significant 
thermal variations of the total magnetization occur.
It is quite apparent from FIG.~\ref{fig5}(a) that a steep decrease in the total magnetization of 
the {\it ferrimagnetic} system to be observed at $D/J_{\rm AB} = -0.45$ can be explained through 
the vigorous thermal excitations $\frac{3}{2} \rightarrow \frac{1}{2}$ of the decorating spins $S_{\rm C}$, which can be clearly seen in the temperature dependence of the sublattice magnetization $m_{\rm C}$. 
On the other hand, 
the opposite thermal excitations $\frac{1}{2} \rightarrow \frac{3}{2}$ are responsible for a rapid increase 
of the sublattice magnetization $m_{\rm C}$ [see for instance the case $D/J_{\rm AB} = -0.55$ shown in FIG.~\ref{fig5}(b)], which in turn causes a gradual temperature-induced increase of the total magnetization. As far as the mixed {\it ferro-ferrimagnetic} system with $J_{\rm AB}>0$ and $J_{\rm AC}<0$ is concerned, the initial decrease in the total magnetization at $D/J_{\rm AB} = -0.45$ can be again attributed to the vigorous 
thermal excitations $\frac{3}{2} \rightarrow \frac{1}{2}$ of the decorating spins $S_{\rm C}$. However, it is easy to observe from FIG.~\ref{fig5}(a) that the thermal excitations of the decorating spins $S_{\rm B}$ overwhelm in the region of moderate temperatures (i.e. the sublattice magnetization $m_{\rm B}$ declines 
more rapidly with increasing temperature than the sublattice magnetization $m_{\rm C}$), what consequently leads to a slight increase of the total magnetization in the mixed ferro-ferrimagnetic system (remember 
that the sublattice magnetization $m_{\rm B}$ is now oriented in opposite to the sublattice magnetization $m_{\rm A}$ and $m_{\rm C}$). At $D/J_{\rm AB} = -0.55$, the total magnetization of the ferro-ferrimagnetic system starts from zero because the sublattice magnetization $m_{\rm B}$ effectively cancels out both the sublattice magnetization $m_{\rm A}$ and $m_{\rm C}$. The observed temperature-induced increase of the 
total magnetization can be mainly related to the thermal excitations $\frac{1}{2} \rightarrow \frac{3}{2}$ 
of the decorating spins $S_{\rm C}$, which are reflected in the temperature-induced increase of the  sublattice magnetization $m_{\rm C}$.

Another notable temperature dependences of the total magnetization occur in the vicinity of the boundary
value $D_{1,0}/J_{\rm AB} = -1$ at which the decorating spins $S_{\rm B}$ change their spin state. 
As a result, the relatively sharp decrease of the sublattice magnetization $m_{\rm B}$ to be observed 
at $D/J_{\rm AB} = -0.9$ leads in the ferrimagnetic system to the relevant decrease of the total 
magnetization [FIG.~\ref{fig4}(a)], while in the mixed ferro-ferrimagnetic system [FIG.~\ref{fig4}(b)] 
it is responsible for an effective increase of the total magnetization since the sublattice magnetization $m_{\rm A}$ and $m_{\rm C}$ are almost completely independent of temperature [FIG.~\ref{fig5}(c)]. Last 
but not least, the most interesting temperature dependences of the total magnetization could be expected 
in the parameter space, where the 2D decorated mixed-spin system should effectively behave as the 'quasi-1D' system. FIG.~\ref{fig5}(d) shows the sublattice magnetization for the particular value of single-ion anisotropy $D/J_{\rm AB} = -1.1$, which is sufficiently strong to force all the integer-valued decorating spins $S_{\rm B}$ towards their 'non-magnetic' state $S^{\rm B} = 0$ at zero temperature. As one can see 
from this figure, the sublattice magnetization $m_{\rm B}$ indeed starts from zero in agreement with our expectations and it is merely the effect of temperature that causes a steady rise of the sublattice magnetization $m_{\rm B}$ on behalf of promoting the $0 \rightarrow 1$ excitations of the decorating 
spins $S_{\rm B}$. Altogether, it might be concluded that the shape of temperature dependence of the total magnetization is in the ferrimagnetic system almost entirely determined by the sublattice magnetization $m_{\rm B}$ (the sublattice magnetization $m_{\rm A}$ and $m_{\rm C}$ almost completely cancel out), while 
in the mixed ferro-ferrimagnetic system the rising magnetization $m_{\rm B}$ lowers the mutual contribution 
of the sublattice magnetization $m_{\rm A}$ and $m_{\rm C}$ to the total magnetization. 

\section{\label{sec:conc}Concluding remarks}

In the present article, the generalized decoration-iteration mapping transformation has been utilized 
to obtain an exact solution of the mixed spin-($S_{\rm A}$, $S_{\rm B}$, $S_{\rm C}$) Ising model on 
the decorated square lattice with the two different kinds of decorating spins $S_{\rm B}$ and  $S_{\rm C}$ placed on its horizontal and vertical bonds, respectively. Within the framework of this exact method, the sought exact solution for the model under investigation has been attained by establishing 
a simple mapping relationship with the corresponding spin-1/2 Ising model on the anisotropic square (rectangular) lattice whose exact solution is known since Onsager's pioneering work \cite{onsa44}. 
In addition, the applied transformation method is of immense practical importance, since this method 
is rather general and it enables further interesting extensions. Actually, it is quite straightforward to extend the applied procedure to account for: 
(i) the interaction between the next-nearest-neighboring spins $S_{\rm A}$; (ii) the multi-spin interaction between the decorating spins and their nearest neighbors; (iii) the biaxial single-ion anisotropy acting on the decorating sites; (iv) other decorated lattices such as decorated honeycomb 
or triangular lattices with (two or even three) different decorating spins; (v) decorated lattices 
with two or more decorating spins per one bond; etc. It is noteworthy, moreover, that the applied method is in principle applicable also to 3D decorated lattices, but unfortunately we cannot present 
an exact solution of the 3D decorated models because of the unknown exact solution of the corresponding spin-1/2 Ising model on the 3D lattice (however, some results with a high numerical accuracy are available even for 3D lattices \cite{oitm03}).

The most interesting result to emerge from the present study consists in providing an exact evidence 
for the spontaneous long-range order, which surprisingly appears in the 2D decorated spin system
in spite of its 'quasi-1D' character. As a matter of fact, we have found a convincing evidence that 
the 2D decorated spin system remains spontaneously long-range ordered even if all the integer-valued decorating spins $S_{\rm B}$ are driven by a sufficiently strong (but not too strong) single-ion anisotropy towards their 'non-magnetic' state $S^{\rm B} = 0$ and the system becomes 'quasi-1D' 
due to the effective splitting into a set of independent mixed spin-$(S_{\rm A}, S_{\rm C})$ chains. 
This finding has obvious relevance to the understanding of the 'quasi-1D' spin systems prone to spontaneous long-range ordering below some critical temperature, which necessarily need not arise 
from interactions establishing 3D connectivity (3D magnetic lattice), but it may represent an 
inherent feature of the 'quasi-1D' system. It is worthwhile to remark that this outstanding feature cannot be found in the decorated spin system with both the half-odd-integer decorating spins $S_{\rm B}$ and $S_{\rm C}$ \cite{cano06b}, while it might be found in the decorated spin system with both 
the integer-valued decorating spins $S_{\rm B}$ and $S_{\rm C}$ provided that they are coupled to 
the vertex spins $S_{\rm A}$ through two different exchange interactions $J_{\rm AB} \neq J_{\rm AC}$, respectively \cite{cano06a}. From this point of view, the decorated spin system with the 
integer and half-odd-integer decorating spins $S_{\rm B}$ and $S_{\rm C}$ exhibits the 
spontaneous ordering related to an appearance of the 'quasi-1D' spin system the most naturally
and hence, this exactly solved model would be of major importance if some experimental realization 
of it would confirm an appearance of the spontaneous ordering (and consequently non-trivial criticality) notwithstanding of its 'quasi-1D' character. 

Although the Ising model description may not be fully realistic for true magnetic materials, it is 
quite reasonable to expect that the exact solution of this simplified model illustrates many important 
vestiges of the real critical behavior. Besides, the exact solution may also bring other valuable 
insights into the thermodynamical properties (magnetization, entropy, specific heat) of real 
magnetic materials without being affected by any crude and/or uncontrollable approximations. 
According to this, the main stimulus for the study of the mixed spin-($S_{\rm A}$, $S_{\rm B}$, $S_{\rm C}$) 
Ising model on the anisotropically decorated square lattice can be viewed in connection with possible experimental realization of this remarkable and rather curious magnetic lattice. It is therefore of 
particular interest to mention that polymeric compounds, which have the decorated square network assembly
as the magnetic lattice, have been rather frequently prepared in an attempt to design novel bimetallic coordination compounds. 
The magnetic structure of the decorated square lattice is indeed peculiarity of two numerous series 
of polymeric coordination compounds with the general formula [Ni(L)$_2$]$_2$[Fe(CN)$_6$]X.nH$_2$O 
\cite{ohba95} and A[M$_{\rm B}$(L)]$_2$ [M$_{\rm A}$(CN)$_6$].nH$_2$O (M$_{\rm A}$ = Fe, Mn, Cr, Co; 
M$_{\rm B}$ = Mn, Fe) \cite{miya95}. In the former series, the magnetic Fe$^{3+}$ ($S_{\rm A} = 1/2$) 
ions reside the square lattice sites and Ni$^{2+}$ ($S_{\rm B} = 1$) ions decorate each its bond, 
while in the latter series the high-spin M$_{\rm B}^{3+}$ ions like Mn$^{3+}$ ($S_{\rm B} = 2$) or 
Fe$^{3+}$ ($S_{\rm B} = 5/2$) occupy the decorating sites and the low-spin M$_{\rm A}^{3+}$ ions 
such as Fe$^{3+}$ ($S_{\rm A} = 1/2$), Mn$^{3+}$ ($S_{\rm A} = 1$), or Cr$^{3+}$ ($S_{\rm A} = 3/2$)
reside the square lattice sites. Even though we are not aware of any trimetallic polymeric compound 
whose network assembly would consist of two different kinds of magnetic metal ions (decorating spins 
$S_{\rm B} \neq S_{\rm C}$) placed on the square net made up by the third magnetic metal ion 
(vertex spin $S_{\rm A}$), the vast number of the bimetallic coordination compounds from 
the aforementioned series gives us hope that a targeted synthesis of such trimetallic compound 
(tailored from two structural derivatives with different decorating magnetic ions) could be 
successfully accomplished in the near future.

\begin{acknowledgments}
This work was supported in part by the Slovak Research and Development Agency under the 
contract Nos. LPP-0107-06 and APVT 20-005204. The financial support provided under the grants 
Nos. VVGS 12/2006 and VEGA 1/2009/05 is also gratefully acknowledged.
\end{acknowledgments}

\end{document}